\documentclass{aa}  
 
\usepackage{graphicx}
\usepackage{txfonts}
\begin{document}
   \title{Magnetic loop emergence within a granule}


   \author{P. G\"{o}m\"{o}ry\inst{1,4},
           C. Beck\inst{2}, 
           H. Balthasar\inst{3},
           J. Ryb\'{a}k\inst{4}, 
           A. Ku\v{c}era\inst{4},
           J. Koza\inst{4},  
           \and
           H. W\"{o}hl\inst{5}  
          }

\institute{IGAM/Kanzelh\"{o}he Observatory, Institute of Physics, 
           Universit\"{a}t Graz, Universit\"{a}tsplatz 5, 8010 
           Graz, Austria \\
              \email{gomory@astro.sk}
           \and
            Instituto de Astrof\'{\i}sica de Canarias (IAC), Via 
            L\'{a}ctea, 38205, La Laguna, Spain
           \and
            Astrophysikalisches Institut Potsdam, An der Sternwarte 
            16, 14482 Potsdam, Germany  
           \and
            Astronomical Institute, Slovak Academy of Sciences, 
            05960 Tatransk\'{a} Lomnica, Slovakia
           \and
            Kiepenheuer-Institut f\"{u}r Sonnenphysik, 
            Sch\"{o}neckstr.\,6, 79104 Freiburg, Germany 
          }

   \date{Received XXX; accepted XXX}

 
  \abstract
   {}
   {We investigate the temporal evolution of magnetic flux emerging 
    within a granule in the quiet-Sun internetwork at disk center.}
   {We combined IR spectropolarimetry of high angular resolution performed 
    in two \ion{Fe}{i} lines at 1565\,nm with speckle-reconstructed G-band 
    imaging. We determined the magnetic field parameters by a LTE inversion 
    of the full Stokes vector using the SIR code, and followed their evolution 
    in time. To interpret the observations, we created a geometrical model 
    of a rising loop in 3D. The relevant parameters of the loop were matched 
    to the observations where possible. We then synthesized spectra from the 
    3D model for a comparison to the observations.}
   {We found signatures of magnetic flux emergence within a growing granule. 
    In the early phases, a horizontal magnetic field with a distinct 
    linear polarization signal dominated the emerging flux. Later on, two 
    patches of opposite circular polarization signal appeared symmetrically 
    on either side of the linear polarization patch, indicating a small 
    loop-like structure. The mean magnetic flux density of this loop was 
    roughly 450\,G, with a total magnetic flux of around 3$\times 10^{17}$\,Mx. 
    During the $\sim$12\,min episode of loop occurrence, the spatial extent 
    of the loop increased from about 1 to 2\,arcsec. The middle part of the 
    appearing feature was blueshifted during its occurrence, supporting the 
    scenario of an emerging loop. There is also clear evidence for the interaction 
    of one loop footpoint with a preexisting magnetic structure of opposite 
    polarity. The temporal evolution of the observed spectra is reproduced to 
    first order by the spectra derived from the geometrical model. During the 
    phase of clearest visibility of the loop in the observations, the observed 
    and synthetic spectra match quantitatively.} 
   {The observed event can be explained as a case of flux emergence in the 
    shape of a small-scale loop. The fast disappearance of the loop at the 
    end could possibly be due to magnetic reconnection.} 

   \titlerunning{Magnetic loop emergence within a granule}
   \authorrunning{G\"{o}m\"{o}ry et al.}
   
   \maketitle
%
\section{Introduction} \label{introduction}

Our knowledge of small-scale magnetic fields located in the 
photosphere and low chromosphere has evolved extremely fast thanks 
to the high spatial resolution observations done with the Hinode 
satellite 
(e.g., Lites et al. \cite{litesetal08}),
and the improvement of the spatial resolution of large ground-based 
solar telescopes achieved by adaptive optics systems
(e.g., von der L{\"u}he et al. \cite{vdlueheetal03}).
The results of such observations tracing the weakest magnetic fields 
can now be compared directly with theoretical models, for example 
on the question whether a local dynamo operates in the solar photosphere 
(Cattaneo \cite{cattaneo99}). 
The observations may also help to refine modern numerical simulations 
of surface magnetoconvection  
(V\"{o}gler et al. \cite{vogleretal05}, 
Schaffenberger et al. \cite{schaffenbergeretal06},  
Stein \& Nordlund \cite{stein_nordlund06}, 
Abbett \cite{abbett07}) 
by providing more stringent boundary conditions on magnetic field 
properties in the solar photosphere. 

Direct measurements of the 3D topology and the temporal evolution of 
magnetic loops associated with small granular and intergranular structures 
are, however, still rare. The measurements require accurate 
spectropolarimetric 2D data of all Stokes parameters with high spatial 
resolution, high signal-to-noise ratio, and high cadence. 
Mart\'{\i}nez Gonz\'{a}lez et al. (\cite{martinezgonzalezetal07}) 
reported observations of the magnetic field vector in the internetwork. 
They showed evidence that small-scale (only 2-6\,arcsec long) low-lying 
loops connect at least 10-12\% of the internetwork magnetic flux measured 
at 1565\,nm.

Centeno et al. (\cite{centenoetal07}, CE07) 
and                  
Orozco Su\'{a}rez et al. (\cite{orozcosuarezetal08}, OR08) 
analyzed the temporal evolution of such small-scale magnetic features. 
They used similar data sets of the \ion{Fe}{i}\,630.15\,nm and 630.25\,nm 
spectral lines from the Hinode/SOT spectropolarimeter
(Lites et al. \cite{litesetal01},
Kosugi et al. \cite{kosugietal07}). 
OR08 
constructed maps of the temporal evolution of linear and circular 
polarization signals, line-of-sight velocity, and the intensities 
of the \ion{Fe}{i}\,630.25\,nm and \ion{Ca}{ii}\,H spectral line. 
They found magnetic signals appearing at the central parts of 
granules, but without a significant linear polarization signal. 
Estimated lifetimes of the events were 20 and 14\,min. They did 
not find any significant coupling of the events to the chromosphere.
CE07 
applied an inversion assuming local thermodynamic equilibrium (LTE) 
to the Hinode spectra to retrieve information on the temporal 
evolution of the magnetic flux and its topology. Similar to  
OR08, 
they also reported the appearance of magnetic signals in the 
central parts of granules, and documented a drift of vertical 
dipoles toward the surrounding intergranular lanes. CE07 found 
a slightly shorter lifetime of only 8 min. Moreover, they found 
that the appearance of horizontal magnetic fields precedes that 
of the vertical fields. They interpreted their results as the 
rise of small-scale magnetic loops through the photosphere, 
although they could not exclude a descending loop.

In this paper, we present a similar analysis of the temporal 
evolution of the appearance and topology changes during a particular 
flux emergence event in the internetwork at the disk center, using 
ground-based observations in the infrared spectral region rendering 
high polarimetric precision and magnetic sensitivity. We compare 
these observations with synthetic spectra from a geometrical model 
of a rising loop to investigate whether the observations can be 
interpreted in such terms.
\section{Observations} \label{observations}

We observed a very quiet region at disk center using the German 
Vacuum Tower Telescope 
(VTT, Schr\"{o}ter et al. \cite{schroteretal85})
located at the Observatorio del Teide (Tenerife, Spain) on June 6, 
2008 between 09:57:50\,-\,11:10:30\,UT.

We used the Kiepenheuer-Institut Adaptive Optics System (KAOS,  
Berkefeld \cite{berkefeld06})
to improve the spatial resolution by a real-time correction of 
wavefront distortions. The uncorrected Fried parameter $r_{\it 0}$ 
reached an average value of about 9\,cm during the whole observing 
run that is sufficient to reach the diffraction limit of the VTT of 
around 0\farcs6 at infrared (IR) wavelengths after the correction.

The light coming from the telescope was separated into visible and 
IR radiation components using a dichroic beam splitter (splitting edge 
at 800 nm). The IR radiation was sent to the Tenerife Infrared Polarimeter 
(TIP 2,
Collados et al. \cite{colladosetal07})
attached to the main spectrograph of the VTT. The visible fraction 
of the radiation was reflected to an optical bench where the following 
instruments were fed in parallel by two additional achromatic beam 
splitters: a G-band broad-band filter (center and {\em FWHM}: 430.6\,nm, 
0.9\,nm), a narrow-band Zeiss Lyot filter for H$\alpha$ ({\em FWHM} 
$\sim$0.025\,nm), and the TESOS 2D spectrometer 
(Kentischer et al. \cite{kentischeretal98}, 
Tritschler et al. \cite{tritschleretal02}).
More information on the instrumental setup can be found also in 
Ku\v{c}era et al. (\cite{kuceraetal08}) or 
Beck et al. (\cite{becketal07}).
%
  \begin{figure}
    \centering
         \resizebox{8.8cm}{!}{\includegraphics{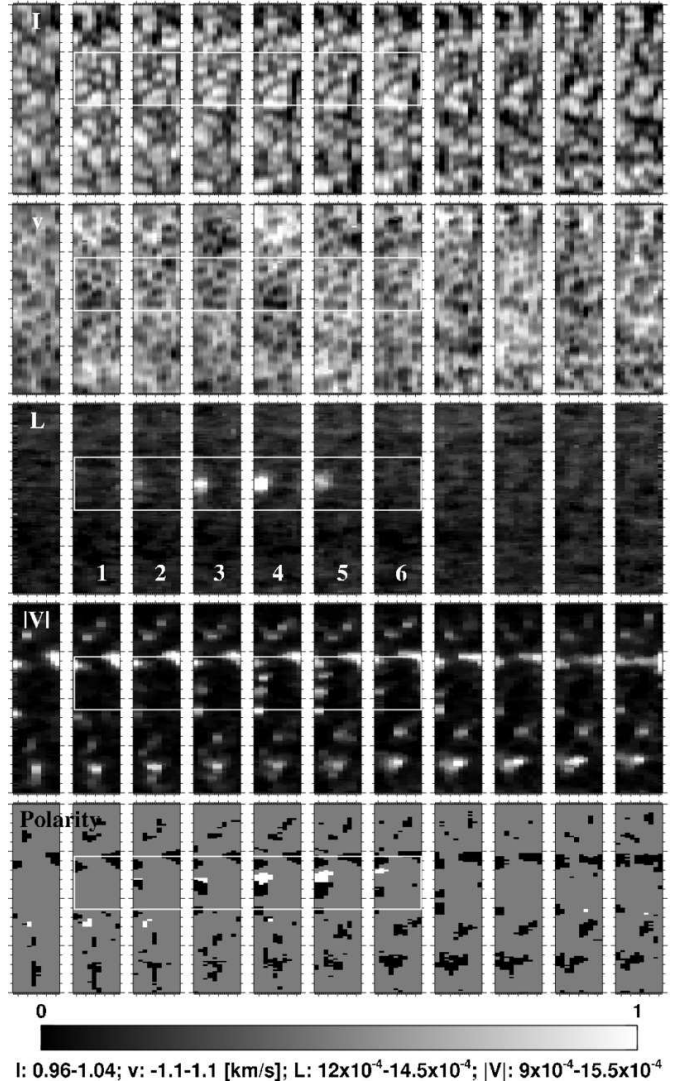}}
         \caption{Overview of the TIP dataset. The panels show from 
                  {\em top to bottom} 2D maps of the continuum 
                  intensity, the Doppler shifts of the \ion{Fe}{i} 
                  1564.85\,nm line (black to mid-gray colors 
                  correspond to the blueshifts, mid-gray to white 
                  colors represent the redshifts), the net linear 
                  polarization, the circular polarization, and the 
                  polarity of the detected magnetic structures. The 
                  size of a single map here is 
                  5$^{\prime\prime}\times$20$^{\prime\prime}$; their 
                  corresponding positions inside the G-band FoV are 
                  marked in Fig.\,\ref{fig_gb_overview}.   
                  }
         \label{fig_tip_overview}
  \end{figure}
%
%
  \begin{figure*}
    \centering
         \resizebox{18.1cm}{!}{\includegraphics{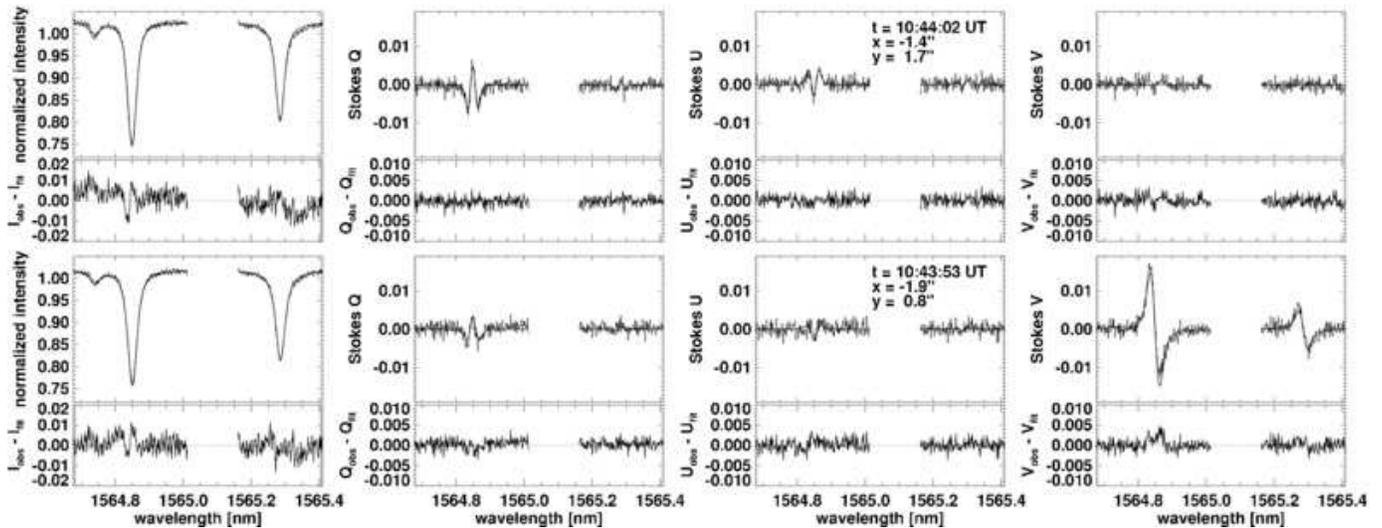}}
         \caption{Examples of the typical Stokes profiles detected in 
                  the middle part ({\em top row}) and southern footpoint 
                  ({\em bottom row}) of the emerging magnetic feature. 
                  {\em Thin lines} show observed profiles, {\em thick 
                  lines} the best-fit profiles of the inversion. Narrow 
                  panels show their residual differences. The 
                  {\em x, y, t} parameters represent spatial positions 
                  of extracted profiles and the times of particular 
                  exposures. 
                  }
         \label{fig_tip_prof}
   \end{figure*}
%

We used the TIP spectropolarimeter with a 0\farcs45 wide and 
77$^{\prime\prime}$ long slit to repeatedly scan a narrow 
5$^{\prime\prime}$ wide area in 10 steps. TIP covered the spectral 
region around the \ion{Fe}{i} lines at 1564.85\,nm and 1565.28\,nm. 
For a more detailed description of these lines see, e.g.,  
Bellot Rubio et al. (\cite{bellotrubioetal04}) 
or
Khomenko \& Collados (\cite{khomenko_collados07}). 
Spatial and spectral sampling of the data were 0\farcs18\,/px and 
14.64\,m\AA/px, respectively. We acquired the four modulation states 
employed by TIP with 250\,ms exposures each. To obtain a higher 
signal-to-noise ratio in the recorded Stokes parameters ({\em I}, 
{\em Q}, {\em U}, {\em V}), we then added 8 sets of the modulation 
states at each scan position, yielding a total integration time of 
8\,s and a cadence of $\sim$10\,s for two consecutive scanning positions. 
The acquisition of the whole 10-step raster thus took about 100\,s 
for every repetition of the scanning.

The G-band images were acquired with a 14-bit PCO 4000 digital camera. 
We set the exposure time to 5\,ms and the cadence to 0.6\,s; the field 
of view (FoV) was 70$^{\prime\prime}$\,$\times$\,100$^{\prime\prime}$, 
and the spatial sampling was 0\farcs08 per pixel. The H$\alpha$ data 
(see, e.g., Ku\v{c}era et al. \cite{kuceraetal08}) 
and the data taken with the TESOS instrument were not used for this 
publication. 
\section{Data reduction and analysis} \label{data_reduction} 

\begin{figure*}
    \centering
         \resizebox{17.85cm}{!}{\includegraphics{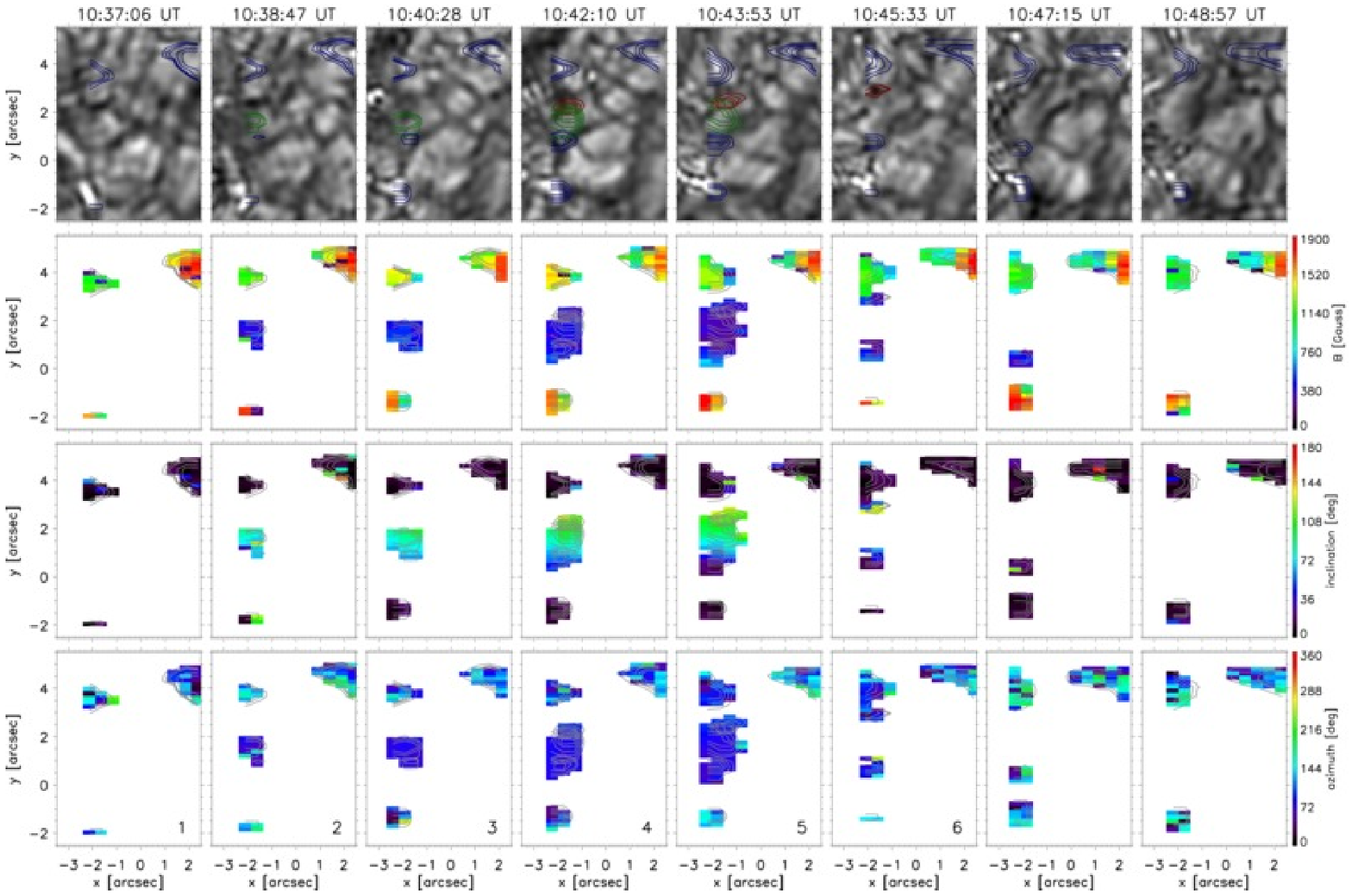}}
         \caption{Temporal evolution during the flux emergence. The 
                  G-band images are shown in the {\em 1$^{st}$ row}. 
                  Areas with enhanced net linear polarization are 
                  marked by {\em green} contours. The {\em blue} and 
                  {\em red} contours represent positive and negative 
                  circular polarization, respectively. The outer contours 
                  enclose regions where linear and circular polarization 
                  are greater than 0.13\% and 0.11\%, respectively, 
                  and the innermost contours represent a 0.16\% level 
                  for both kinds of polarization. 
                  The other rows show the magnetic flux density 
                  {\em (2$^{nd}$ row)}, inclination {\em (3$^{rd}$ row)},  
                  and azimuth {\em (4$^{th}$ row)}. The {\em gray contours} 
                  mark the same polarization levels as before. 
                  The times given on {\em top} correspond to moments 
                  when TIP passed across the occurring feature during the 
                  scanning.
                  }
         \label{fig_gband_magparam}
  \end{figure*}

We photometrically corrected all data by subtracting the dark current 
and applying flat-field frames. To compensate for the instrumental 
polarization, we applied the correction described by 
Schlichenmaier \& Collados (\cite{schlichenmaier_collados02})
to the spectra of TIP, using a data set obtained with the IR 
instrument calibration unit of the VTT. Then, for a complete 
correction, the coelostat configuration of the telescope was 
modeled according to 
Collados (\cite{collados99}) and  
Beck et al. (\cite{becketal05}).
We finally estimated statistically the residual crosstalk following 
Collados (\cite{collados03}). 
An overview of the relevant section of the dataset obtained with 
TIP is shown in 
Fig.\,\ref{fig_tip_overview}. 
It shows from {\em top to bottom} the continuum intensity $I_{c}$, 
the line-core Doppler shifts of the \ion{Fe}{i} 1564.8\,nm line,  
the total linear polarization $L = \int (Q^2 + U^2)^{1/2}/ I_{c} d\lambda$, 
the total circular polarization $|V| = \int |V|/I_{c} d\lambda$, 
and the magnetic field polarity derived from the order of minimum 
and maximum of the $V$ signal. The region of the flux emergence is 
marked with {\em white rectangles}; the repetitions of the scanning 
marked with the consecutive numbers 1 to 6 are used for comparison 
with the geometrical loop model later on 
(see also Fig.\,\ref{fig_gband_magparam}).

To derive the magnetic field parameters, we used the SIR code (Stokes 
Inversion based on Response functions, 
Ruiz Cobo \& del Toro Iniesta \cite{cobo_iniesta92})
for a simultaneous inversion of the Stokes profiles of \ion{Fe}{i} 
lines at 1565\,nm in a similar way as described by 
Bellot Rubio et al. (\cite{bellotrubioetal04}).
We used a two-component model of the solar atmosphere with a magnetic 
and a field-free component that allows us to derive accurate parameters 
also in the case of spatially unresolved magnetic fields. All magnetic 
parameters were kept constant with height during the inversion. To 
demonstrate the quality of the performed inversion, we display the 
typical Stokes profiles detected in the middle part and in the southern 
footpoint of the emerging feature in 
Fig.\,\ref{fig_tip_prof}
together with the corresponding best-fit profiles determined by SIR, 
and their residual differences. The {\em rms} noise in continuum windows 
of the observed Stokes profiles shown in 
Fig.\,\ref{fig_tip_prof}
was $3.5 \times 10^{-3}$, $1.1 \times 10^{-3}$, $1.1 \times 10^{-3}$, 
and $1 \times 10^{-3}$ for the middle part and $3.7 \times 10^{-3}$, 
$1.1 \times 10^{-3}$, $1.5 \times 10^{-3}$, and $8 \times 10^{-4}$ for 
the southern footpoint in the {\em I}, {\em Q}, {\em U}, and {\em V} 
Stokes profiles, respectively.

\begin{figure*}
    \centering
        \resizebox{16.6cm}{!}{\includegraphics{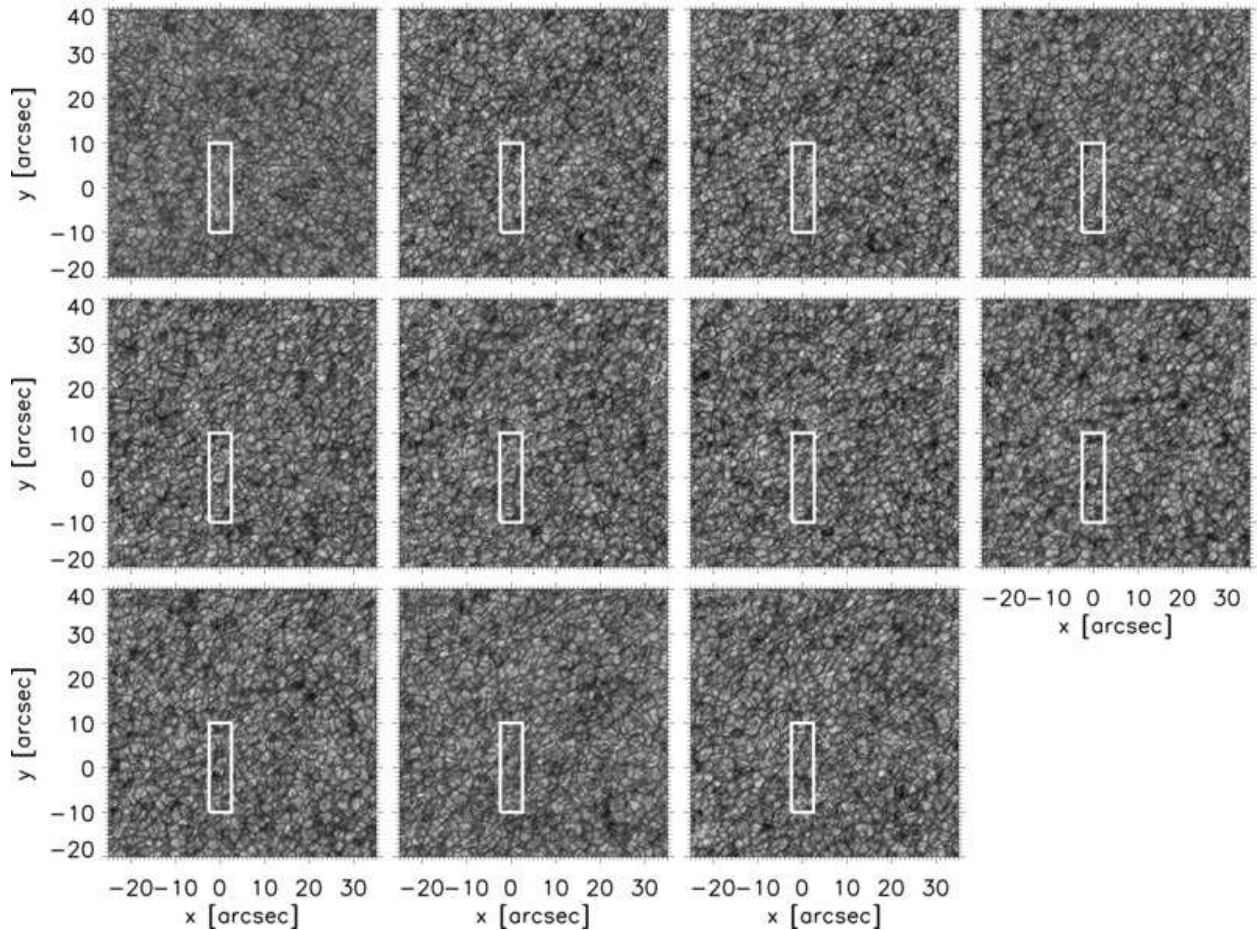}}
        \caption{Sample of the speckle-reconstructed G-band images 
                 showing the surroundings of the observed region. 
                 The locations of the TIP scans shown in 
                 Fig.\,\ref{fig_tip_overview} are marked by 
                 {\em white rectangles}. The zero point of the 
                 coordinate system is shifted 55\,arcsec from the 
                 disk center.           
                 }
         \label{fig_gb_overview}
  \end{figure*}

We used the Kiepenheuer-Institut Speckle Interferometry Package (KISIP) 
code
(W\"{o}ger et al. \cite{wogeretal08}) 
for the speckle reconstruction of the G-band data. A burst of 100 images 
was selected for each reconstruction leading to a temporal cadence of 
63\,s. The maximal time difference between a TIP scan position and the 
G-band reconstruction closest in time is thus around 30 seconds. A 
sample of the acquired images is given in 
Fig.\,\ref{fig_gb_overview}.

Because of the importance of a precise co-alignment of all data, 
we took additional grid target data with all involved instruments. 
Using these data, we determined the rotation between the TIP 
spectropolarimetric rasters and the G-band images. The overall 
shift of these datasets, which also includes the effects 
of differential refraction, was then derived with sub-arcsec 
precision using a cross-correlation technique. At the end, we compared 
the co-spatiality of some G-band bright points with enhanced Stokes 
{\em V} signals as an independent check of the applied co-alignment 
method.
%
%
%
%

\section{Emerging loop in observations and inversion} \label{results_obs}

The temporal evolution of the polarization signals in the small 
photospheric internetwork area under study, and the magnetic field 
parameters (i.e., magnetic flux density, field inclination, and 
field azimuth) derived by the inversion are shown magnified in  
Fig.\,\ref{fig_gband_magparam}. 
The first row displays the speckle-reconstructed G-band images. 
The superimposed contours mark areas where a significant polarization 
signal above the noise level was detected. The {\em green} contours 
mark areas of significant net linear polarization. The {\em red} and 
{\em blue} contours represent negative and positive circular polarization, 
respectively, given as $CP = \int (|V| \times polarity)/I_{c}d\lambda$.  

In the first snapshot taken at 10:37:06\,UT, three areas of 
positive circular polarization are visible. These elements 
represent the preexisting magnetic field seen in the data set 
well before, during, and after the time period investigated here. 
The second snapshot (10:38:47\,UT) shows a new, predominantly 
linearly polarized signal at (x,y)=($-2^{\prime\prime}$, 2$^{\prime\prime}$), 
emerging at the location of a granule. In the next minutes, 
the topology of this feature changes significantly. It expands 
rapidly, but mostly within the granular boundaries. The next 
snapshots taken at 10:42:10\,UT and 10:43:53\,UT already show 
areas of opposite circular polarization signal occurring 
symmetrically on both sides of the region with enhanced linear 
polarization signal. The linear and circular polarization patches 
partly overlap at these moments. This configuration could 
correspond to a small magnetic loop-like structure whose footpoints 
are co-spatial to the areas of opposite circular polarization. The 
distance of the potential footpoints is roughly 1-1.5$^{\prime\prime}$. 
A slight downward deformation of the preexisting positive circular 
polarization patch at ($-2^{\prime\prime}$, 4$^{\prime\prime}$), 
clearly already visible at 10:43:53\,UT, suggests the onset of an 
interaction of the emerging negative polarization patch with this 
feature. Later on, at 10:45:33\,UT, the sudden disappearance of the 
linear polarization signal is followed by a weakening of both circular 
polarization patches which have a distance of about 2$^{\prime\prime}$ 
at this moment. Finally, at 10:47:15\,UT  and subsequently at 
10:48:57\,UT, the remnants of the circularly polarized features have 
disappeared completely, and there is no further evidence of their 
re-occurrence later on in the data set. The whole lifetime of the 
detected magnetic feature thus was less than 12 minutes. 

The overall course of the event seems to be rather asymmetric in time. 
Figure\,\ref{fig_gband_magparam} 
indicates a gradual growth of the area of enhanced net linear 
polarization and the polarization amplitude itself in the four-panel 
sequence between 10:38:47\,UT and 10:43:53\,UT, followed by a sudden 
disappearance in the next panel. The flux emergence leaves no detectable 
brightness patterns on the underlying granule, which grows steadily in 
size. The emergent magnetic flux also does not influence the appearance 
and brightness of G-band bright points in the vicinity of the granule 
significantly; most of them stay visible at about the same location for 
the whole time span of the flux emergence.

\begin{figure}
    \centering
         \includegraphics[width=8.2cm]{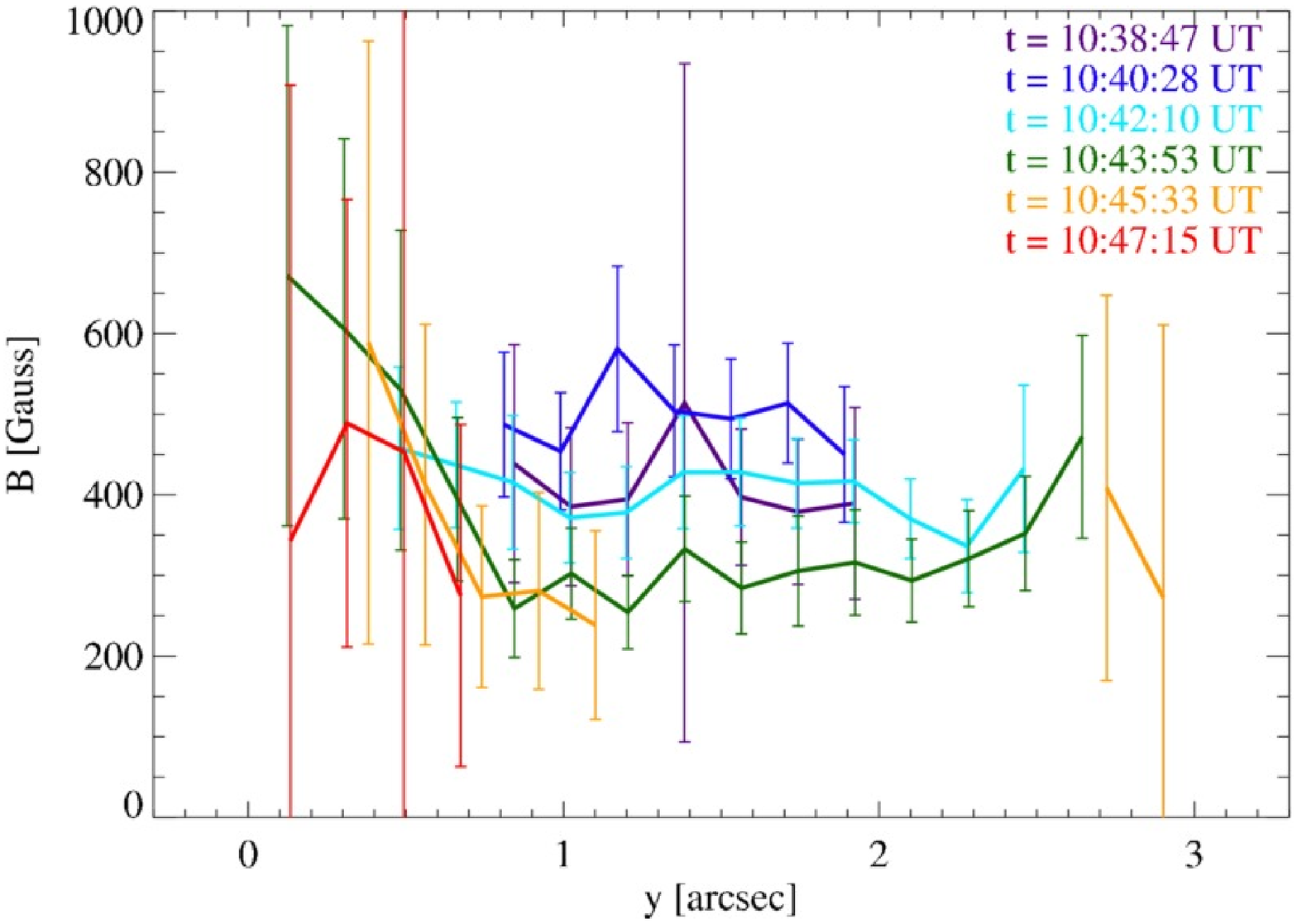}\\
         \includegraphics[bb= 10 -20 575 378, width=8.2cm]{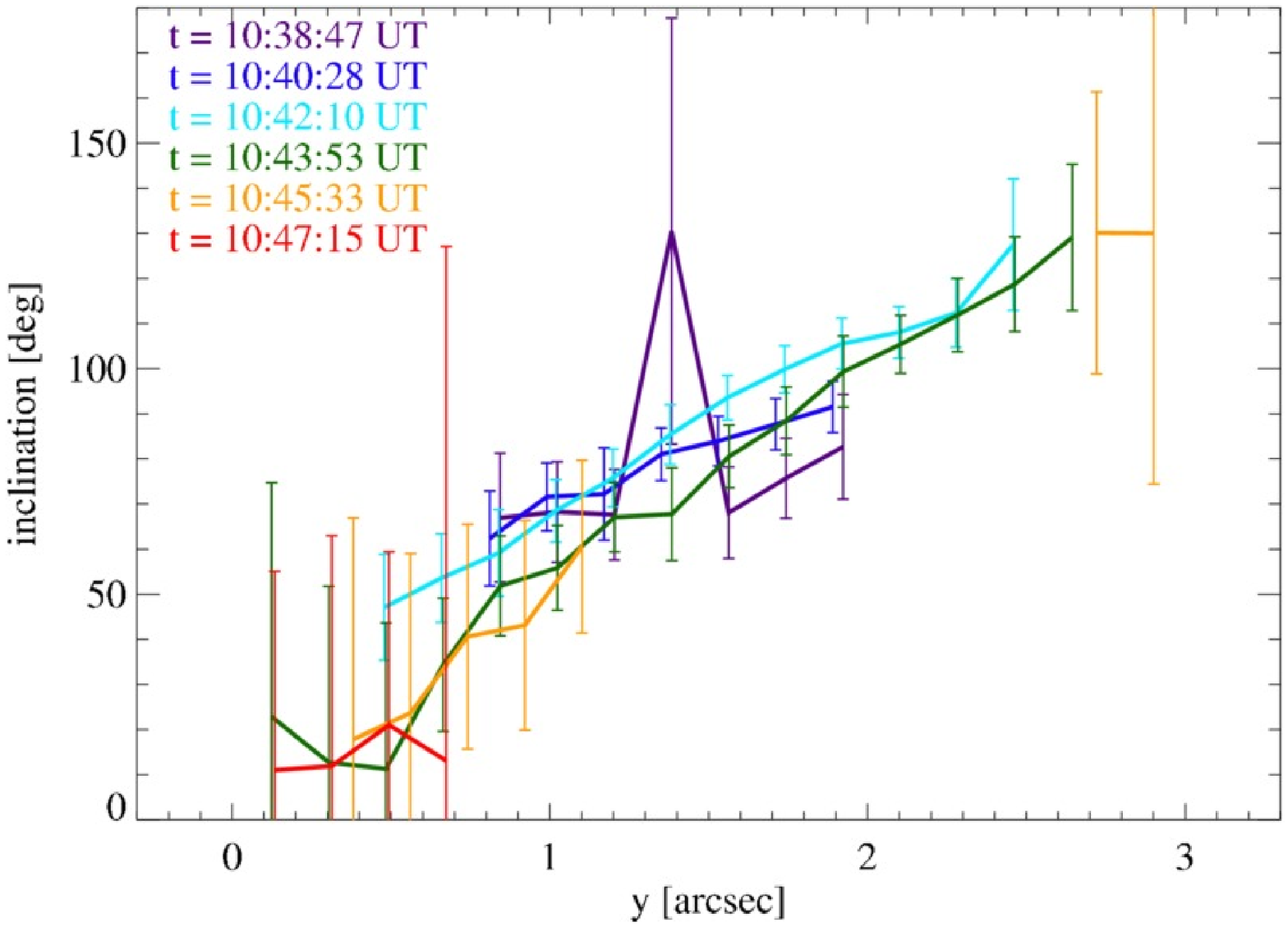}
         \caption{Cuts along the emerging feature showing the 
                  temporal evolution of its magnetic flux density 
                  ({\em top}) and inclination ({\em bottom}).
                  } 
         \label{fig_b_i_evol}
   \end{figure}

The second row of 
Fig.\,\ref{fig_gband_magparam} 
shows the temporal evolution of the magnetic flux density. Compared 
with the preexisting ambient magnetism of kG strength, the flux 
density of the emerging magnetic feature is much weaker and reaches 
a mean value of about 450\,G throughout its 12-min period of occurrence. 
The temporal evolution of this parameter on spatial cuts connecting 
the two footpoints is also shown in the {\em top panel} of
Fig.\,\ref{fig_b_i_evol}.
The displayed error bars correspond to the error estimate of the SIR 
code. Moreover, using the inversion results on the magnetic flux density 
and the magnetic filling fraction, we estimated the total magnetic 
flux seen in each of the two footpoints to be around 3$\times 10^{17}$\,Mx. 
However, the inferred value is very probably affected by the limited spatial 
resolution of the TIP instrument. Therefore, the calculated total magnetic 
flux likely represents only a lower limit. 

The third row of 
Fig.\,\ref{fig_gband_magparam} 
displays the temporal evolution of the magnetic field inclination 
which indicates significant differences between the preexisting 
ambient field and the emerging one. While the former is vertical 
and stays almost constant in time, the latter shows a much more 
complex evolution. In its early stages, the inclination angles are 
close to 90$^\circ$, corresponding to a horizontal magnetic field (see 
also the {\em bottom panel} of
Fig.\,\ref{fig_b_i_evol}).
Later on, the inclination on the edges of the feature becomes more 
vertical. At 10:43:53\,UT, the emerging feature has a clear loop-like 
structure with two opposite, almost vertical polarities connected by 
the horizontal field. The trend for a decrease of the inclination 
continues even after the linear polarization signal has vanished; 
the remaining tracers of the loop footpoints become still more vertical 
and also separate further (see also the lower panel of 
Fig.\,\ref{fig_b_i_evol}).
In the final stage, the upper leg meets with a preexisting magnetic 
structure of opposite polarity. Reconnection of these two magnetic 
elements may thus explain the fast disappearance of the polarization 
signal at this location.  

The bottom row of 
Fig.\,\ref{fig_gband_magparam} 
shows the magnetic azimuth angle. The emerging magnetic structure has 
an almost constant azimuth of $\sim$75$^\circ$ during the whole time 
of its occurrence that is consistent with a rigid magnetic dipole. 
This again supports the proposed model of a magnetic loop. Due to the 
180$^\circ$ azimuth ambiguity, we cannot resolve without additional 
information whether the loop has an $\cap$-like shape and is emerging, 
or an $\cup$-like shape and is submerging 
(CE07). 

\begin{figure}
          \resizebox{8.2cm}{!}{\includegraphics{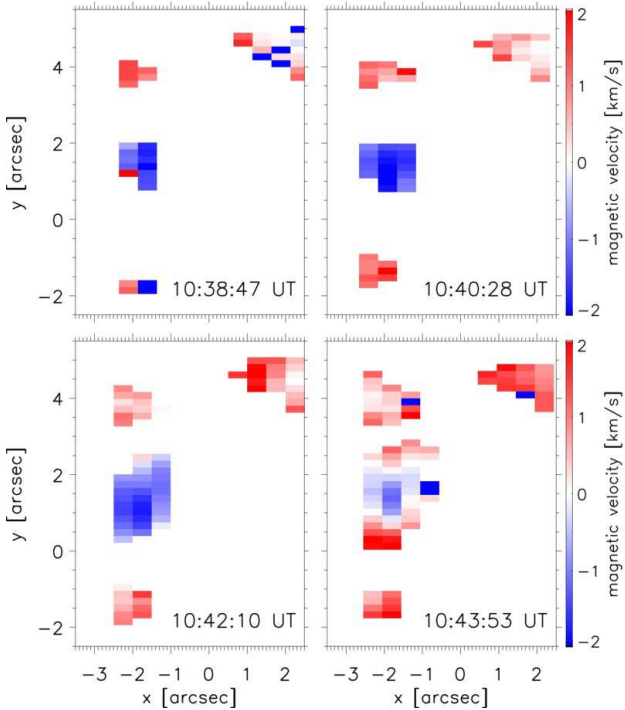}}
          \caption{2D maps showing temporal evolution of the 
                   'magnetic' Doppler shifts. The displayed values 
                   are cut at $\pm$2\,km\,s$^{-1}$.
                  }
          \label{fig_magnetic_vel} 
   \end{figure}

To resolve the mentioned ambiguity between an emerging or submerging 
structure, we analyzed the Doppler shifts of the magnetic inversion 
component. The most relevant maps, showing the phase of the appearance 
of the structure under study, are displayed in 
Fig.\,\ref{fig_magnetic_vel}.
We found that the middle part of the appearing structure was blueshifted 
all the time of its occurrence; the velocity reached up to $-2$\,km\,s$^{-1}$. 
This suggests an emergence of a magnetic loop from the lower atmospheric 
layers. Only later, when the footpoints of the loop become clearly 
visible, we could detect redshifts at the locations of the footpoints, 
corresponding to downflows of up to +2\,km\,s$^{-1}$. This could signify 
that the material in the loop is draining down along the field lines, 
which is in agreement with the generic loop emergence scenarios. The 
surrounding magnetic structures are on the other hand predominantly 
redshifted during the whole displayed time interval. 

Based on the observational findings discussed above, we conclude that 
the temporal evolution, the appearance of the linear and circular polarization 
signal, and the corresponding parameters of the magnetic field together 
with the 'magnetic' Doppler shifts derived by the inversion all suggest 
that the observed flux emergence takes place in the shape of a small loop 
that rises through the photosphere.
%
%
\section{Emerging loop in semiempirical simulation}  \label{simulation_of_loop}
\subsection{Geometrical model} \label{geometrical_model}
To investigate whether the observed event can be explained by flux 
emergence in the shape of a rising loop, we modeled an $\bigcap$-shaped 
loop with the basic geometric parameters as in the observations, 
and synthesized spectra of the IR lines for it. The loop was placed 
in a 3D box of around 3000\,km$\times$1000\,km$\times$500\,km, with 
a spatial sampling of 29\,km in $x$ and $y$, and 1\,km in the $z$ 
direction.

The loop parameters derived from the observations were a horizontal 
separation speed of its footpoints of 2\,km\,s$^{-1}$, a horizontal 
extent of around 600\,km at the beginning, and a life time of around 
600\,s, sampled on 6 maps of the repeated scanning. We then created 
a loop connecting two footpoints with a distance of 
$L(k) = $600\,km\,+\,$k\times 180$\,km $(k=1,2,...,6)$ by prescribing 
a 2$^{nd}$ order polynomial of the shape
%
  \begin{equation}
      z(x) = h_{apex} - \frac{h_{apex}}{(\alpha\times L/2)^2} x^2 \, ,
      \label{eq_zx}
  \end{equation}
%
where $\alpha = 0.6$ and $h_{apex}$ is the location of the apex of 
the loop. The value of $\alpha$ influences the shape of the loop 
by stretching or compressing it. It was also used to take into 
account that the height scale considered for the later conversion 
to optical depth extended beyond 0\,km to $z = -86.2$\,km, whereas 
the cross-section seen in the spectra rather comes from $z>0$ km. 
The loop parameter $h_{apex}$ was determined by trial-and-error to 
match the temporal evolution in the observations to 
$h_{apex}(k) = -65$\,km + $k\times$100\,km, corresponding to a rise 
speed of around 1\,km\,s$^{-1}$.
%
%

\begin{figure}
        \resizebox{9.0cm}{!}{\includegraphics{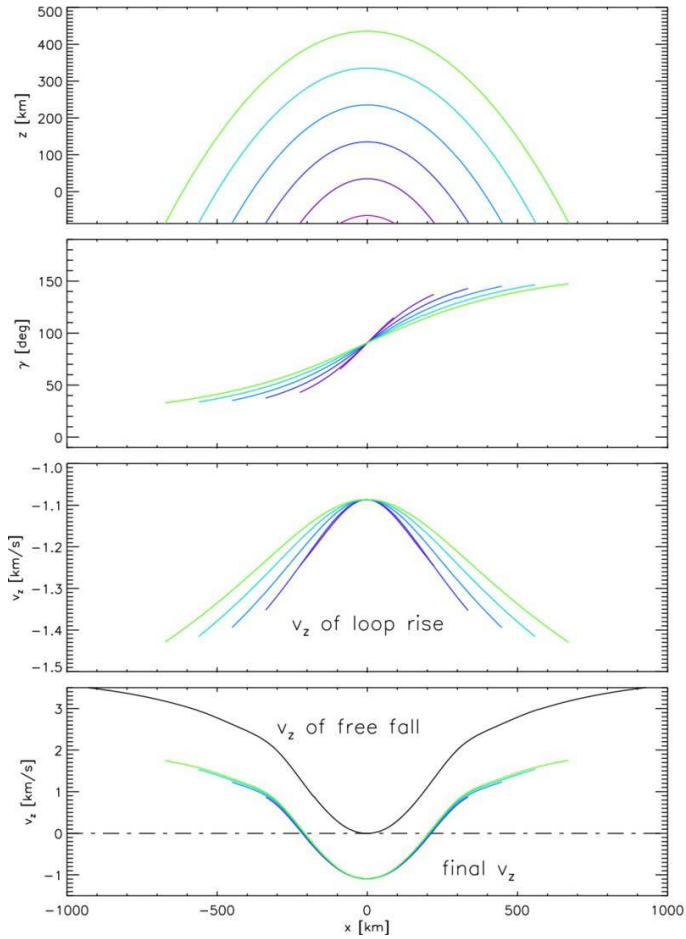}}
        \caption{Temporal development of the simulated loop in six 
                 time steps. From {\em top to bottom}: Central loop 
                 axis, magnetic field inclination, LOS velocity 
                 component $v_z$ of the loop rise, free fall, 
                 and their addition to the final $v_z$.
                 }
        \label{loop_geom}
  \end{figure}

Figure\,\ref{loop_geom} 
shows the resulting shape for the central axis of the loop on the 
six steps on the {\em top}. The inclination of the magnetic field to 
a vertical LOS ({\em 2$^{nd}$ panel}) was taken from the derivative of 
the location of the central axis. We determined an appropriate LOS 
velocity of the material inside the loop from two different 
contributions, the rise of the loop itself (blueshift in the middle) 
and a downflow along the field lines (redshift at the footpoints). 

\begin{figure}
        \centerline{\resizebox{9.1cm}{!}{\includegraphics{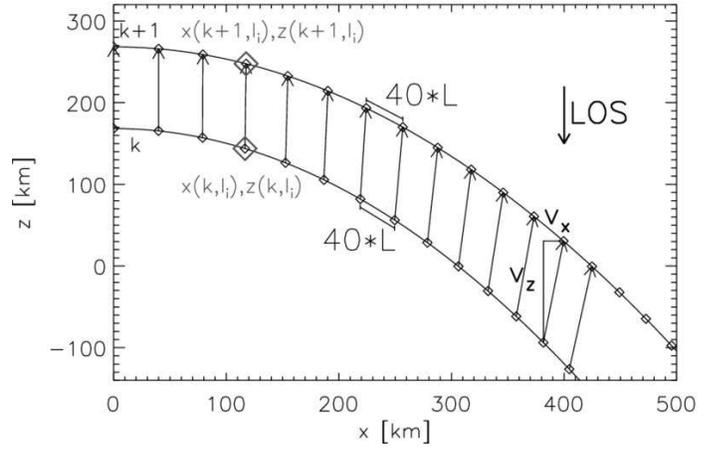}}} 
        \caption{Calculation of velocity components ($v_x,v_z$) 
                 from the location of the central axis in the 
                 steps $k$ and $k+1$.
                 }
        \label{velo_calc}
  \end{figure}

The velocity component induced by the rise of the loop was determined
from the vertical motion of the loop as defined by 
Eq.\,(\ref{eq_zx}).
We parametrized the loop for every step $k$ by equidistant pieces 
$l_i$ of length $L=1$\,km along its axis ($i=0...n_{max}$, where 
$n_{max}$ was set by where the loop reached $z=-1000$\,km), starting 
at the apex at each step, to obtain the positions $x(k,l_i), z(k,l_i)$ 
in the x-z plane 
(see Fig.\,\ref{velo_calc}). 
The velocity component for a vertical LOS at the position $x(k+1,l_i)$ 
on the abscissa was then determined as 
$v_z(x(k+1,l_i)) = (z(k+1,l_i)-z(k,l_i))/dt$, where $dt$ corresponds to 
the $\sim$100\,s cadence of the TIP maps. To obtain $v_z$ for the evenly
spaced grid points used in the spectral synthesis, the velocity values
$v_z(x(k+1,l_i))$ were then interpolated to equidistant 29\,km-spacing 
in $x$. The velocity ranges from $-1$\,km\,s$^{-1}$ at the apex to slightly 
higher velocities of about $-1.4$\,km\,s$^{-1}$ near the footpoints due to 
the additional lateral motion (see {\em 3$^{rd}$ panel} of 
Fig.\,\ref{loop_geom}). 
For the downflow along the field lines, we assumed a steady flow in 
free-fall conditions along the loop geometry of step $k$=6, similar to 
Mein et al. (\cite{meinetal96}) or 
Georgakilas et al. (\cite{georgakilasetal90}, their Eq.~(11)). 
The free-fall condition, however, produces supersonic flows of 
$v\gg$7\,km\,s$^{-1}$ near the footpoints. We thus have set a limit for 
the maximum allowed absolute velocity of 4\,km\,s$^{-1}$, which leads 
to around +2\,km\,s$^{-1}$ vertical velocity near the footpoints 
({\em bottom panel} of 
Fig.\,\ref{loop_geom}).
\subsection{Magnetic and thermodynamical properties} \label{mag_and_therm_prop}
We modeled the loop as a circular flux tube around the central axis 
by using a field strength $B$ with 
%
  \begin{equation}
        B(r) = 500 {\rm \,G} \times \exp\left(\frac{-r^2}{2\sigma^2}\right)\, , 
        \label{eq_b}
  \end{equation}
%
where $r$ measures the distance from the central axis and $\sigma$ 
was set to 100\,km. 

\begin{figure}
         \centerline{\resizebox{8.8cm}{!}{\includegraphics{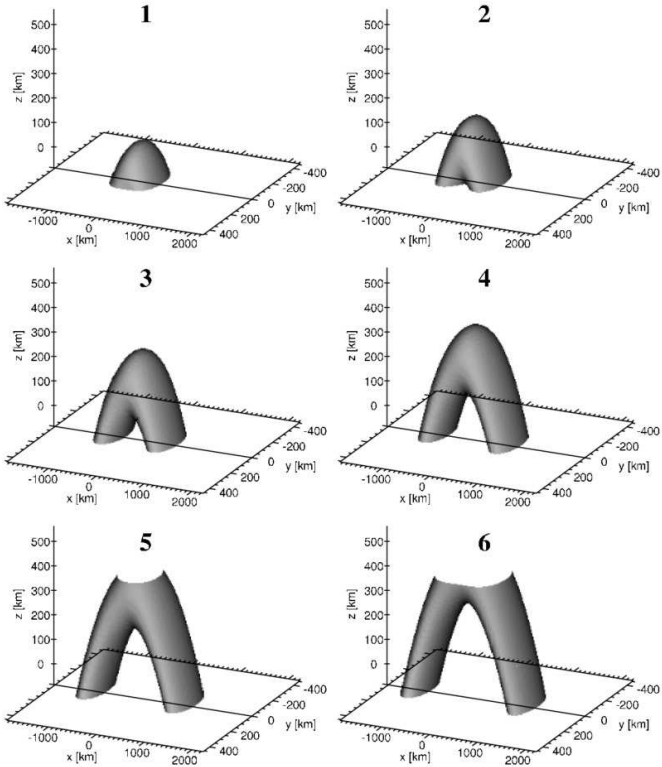}}}
         \caption{Magnetic field topology during the flux emergence 
                  simulation. 
                  } 
         \label{3d_geom} 
  \end{figure}

This yields a flux tube with around 240\,km {\em FWHM} and an effective 
radius of 300\,km, at which distance the field strength drops to 
virtually zero. Integration of
Eq.\,(\ref{eq_b}) 
over the radius yields a total magnetic flux of around 2.7$\times 10^{17}$\,Mx, 
which is slightly below the magnetic flux value derived from the 
inversion results. 
Figure\,\ref{3d_geom} 
displays the final geometry by showing isosurfaces of 100\,G field 
strength for the six steps of the flux emergence. At the first step, 
most of the loop is below the formation height of the spectral lines 
considered, and thus induces only a small polarization signal in the 
spectra. 

We used the tabulated values of the HSRA model 
(Gingerich et al. \cite{gingerichetal71})
to define the thermodynamic properties like temperature, gas density, 
or electron pressure needed for the spectral synthesis. We considered 
an optical depth range from log$\tau$ = +1.4 to around $-3$, which 
covers the formation heights of the IR lines at 1.56\,micron 
(Cabrera Solana et al. \cite{cabrera_solanaetal05}).
We did not consider the effect of the presence of the loop on the 
stratifications (or vice versa) by enforcing hydrostatic equilibrium. 

\begin{figure}[hb!] 
          \centerline{\resizebox{8.8cm}{!}{\includegraphics{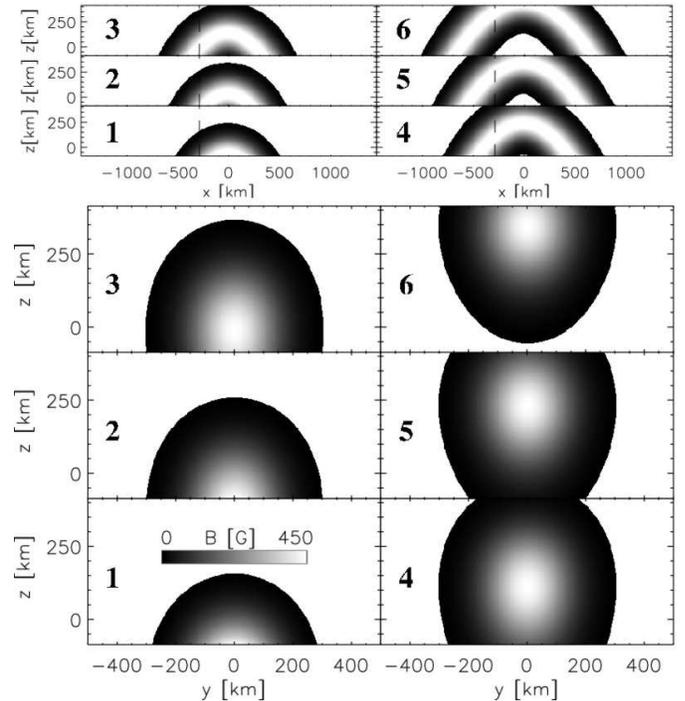}}}
          \caption{Magnetic field strength in side view in the 
                   $x-z$ plane ({\em top}), and cut in the $y-z$ 
                   plane across one of the footpoints ({\em bottom}). 
                   The location of the cut is marked by a {\em 
                   dash-dotted vertical} line in the upper plot.
                   } 
          \label{2d_view} 
   \end{figure}

To determine the magnetic field stratifications for a vertical 
LOS from the 3D topology, we used cuts through the 3D box along 
the $z$-axis. The field strength $B(x,y,z)$ can then be derived 
from the distance of the point $(x,y,z)$ from the closest piece 
of the central axis using 
Eq.\,(\ref{eq_b}). 
The stratifications of $B$ can differ strongly from a Gaussian 
shape, depending on where along the loop the cut is located; near 
the footpoints, one obtains a Gaussian shape stretched in $z$, 
whereas close to the apex the Gaussian shape is preserved 
(Fig.\,\ref{2d_view}).
%
%
%
\subsection{Spectral synthesis and degradation of spatial resolution} \label{spec_synthesis}
Using the derived stratifications for the thermodynamic variables 
and the magnetic field strength, we then synthesized spectra of 
the three spectral lines \ion{Fe}{i}\,1564.8\,nm, a weak blend 
of \ion{Fe}{i} at 1564.7\,nm, and \ion{Fe}{i}\,1565.2\,nm with 
the SIR code. The simulated spectra differ strongly from the 
observations in spatial resolution. To improve the match to the 
observations, we degraded the spatial resolution by a convolution 
with a 2D Gaussian kernel of 0\farcs56 {\em FWHM}, roughly 
corresponding to the diffraction limit of the VTT at IR wavelengths. 
We then calculated 2D maps of total linear and circular polarization 
$L$ and $|V|$, respectively, and the polarity of the Stokes $V$ 
signal analogously to the observations. 

Figure\,\ref{2d_maps} 
shows these maps for both the original and the spatially degraded 
simulation, and the corresponding section from the observation, 
for a visualization of the effects of spatial resolution and spatial 
sampling. The display ranges of each quantity in 
Fig.\,\ref{2d_maps} 
are given in 
Table\,\ref{disp_ranges_f11}. 
For the observations, the value of the wavelength-integrated quantities 
cannot reach zero because of the noise. With the assumption that the 
lowermost value in the observations corresponds to a pure noise 
signal, the dynamic range of total linear polarization $L$ in the 
observations matches fairly well that of the convolved simulation, 
whereas for total circular polarization $|V|$ it is significantly smaller. 
The spatial smearing is sufficient to reduce the circular polarization 
signal in the first step to values below the threshold used for the 
observations (compare {\em 1$^{st}$, 7$^{th}$}, and {\em 13$^{th}$ column}). 
The elongated loop structure seen in $L$ of the full-resolution spectra 
({\em top left}) is lost after the spatial degradation; the strong linear 
polarization signal appears there in the shape of an only slightly elongated 
patch. The two polarities in the degraded synthetic polarity map are seen 
to touch each other persistently up to step 3, similar to the observed 
behavior. The effect of applying the Gaussian kernel is to first order 
already equivalent to sampling the degraded data with the pixel size of 
the observations ({\em white rectangle} in the bottom right panel, 
0\farcs45$\times$0\farcs17 pixel). We thus skipped an additional re-sampling 
with the TIP pixel size for the simulation spectra shown in the following.
%
%
%
\subsection{Direct comparison of observed and synthetic spectra} \label{obs_synt_comparison}
\begin{figure}
          \centerline{\resizebox{8.8cm}{!}{\includegraphics{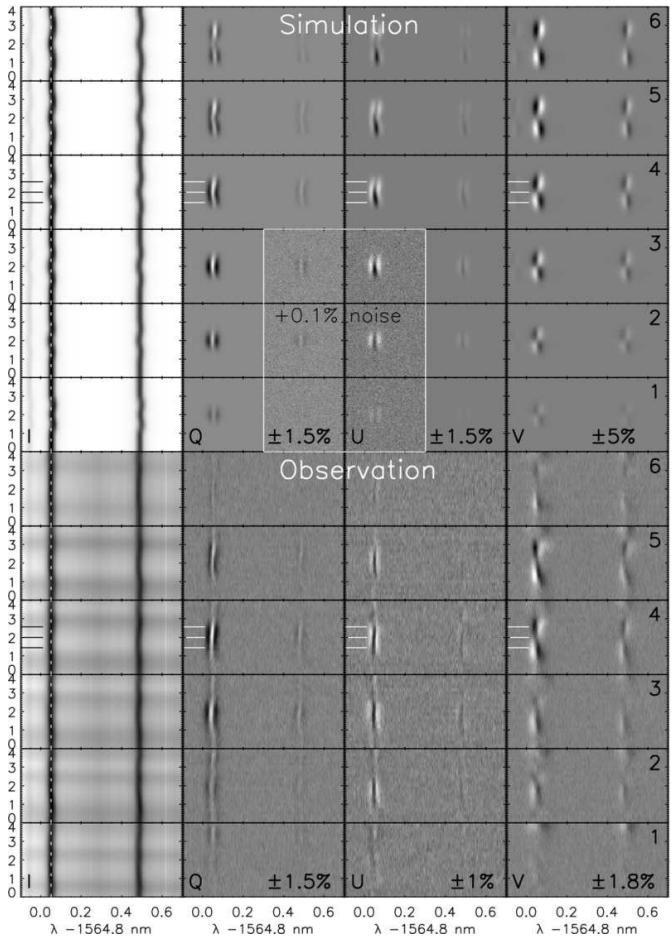}}}
          \caption{Comparison of observed ({\em bottom}) and 
                   simulated spectra ({\em top}). {\em Left to 
                   right}: Stokes $IQUV$. {\em Bottom to top}: 
                   Time steps 1 to 6. The display ranges for 
                   $QUV$ are given at {\em bottom right} in 
                   each panel. The {\em vertical dashed line} 
                   in Stokes $I$ denotes the rest wavelength 
                   of the 1564.8\,nm line. The short {\em horizontal 
                   bars} at the left in $IQUV$ of step 4 denote 
                   the location of the profiles shown in 
                   Fig.\,\ref{fig_profiles}.
                   } 
          \label{spec_comp}
   \end{figure}

For a direct comparison of observed and synthetic spectra, we took 
cuts along the central axis of the feature in the simulation. For 
the observations, we chose the slit spectra of a scan step near the 
beginning of each repeated map, since the structure is roughly oriented 
along the slit. 
Figure\,\ref{spec_comp} 
shows the $IQUV$ spectra for these six cuts. The observed spectra 
({\em at bottom}) show some features that are naturally missing in the 
simulation ({\em at top}): the intensity and velocity pattern of the 
granulation, and additional polarization signal at the upper end of the 
cut corresponding to the preexisting patch of network fields 
(y $\sim 4^{\prime\prime}$). For a comparison of the polarization level in 
the simulation and observation, we added an {\em rms} noise of $1 \times 10^{-3}$ 
to the simulation spectra of Stokes $Q$ of the 1565.2\,nm line and Stokes $U$ 
of the 1564.8\,nm line in the first three steps ({\em white rectangle} at top 
middle). The noise level is sufficient to suppress the polarization signal 
of the less Zeeman-sensitive line at 1565.2\,nm as in the observed spectra. 
The maximal linear polarization amplitudes of the simulation and observation also 
roughly match (up to 1.5\% of $I_c$ for $Q$ and $U$). The Stokes $V$ amplitude 
of the simulation exceeds the observations for the steps 5 and 6, where in 
the observations the cancellation with the preexisting network patch may 
already have happened. The observed trend of polarization signal with time is 
reproduced in the simulation: a monotonical increase of $V$ amplitude until step 
5, increase and decrease for $Q$ and $U$ with maximal amplitude at step 4. 
The observed spectra still show a faint $Q$ and $U$ signal in step 6, only 
seen in the 1564.8\,nm line. Near the footpoints, the observed Stokes $V$ 
signal has a drop-like shape, with a decrease of the splitting and a displacement 
to the blue when moving towards the opposite polarity that also shows up in 
the simulation spectra. The observed Stokes $Q$ signal in step 3 has the 
same curved shape in {\em y} as the synthetic spectra of steps 3 and 4, 
indicating the blueshift in the middle and the redshift near the footpoints. 
%
%
%
%
%
  \begin{figure}
         \centerline{\resizebox{8.8cm}{!}{\includegraphics{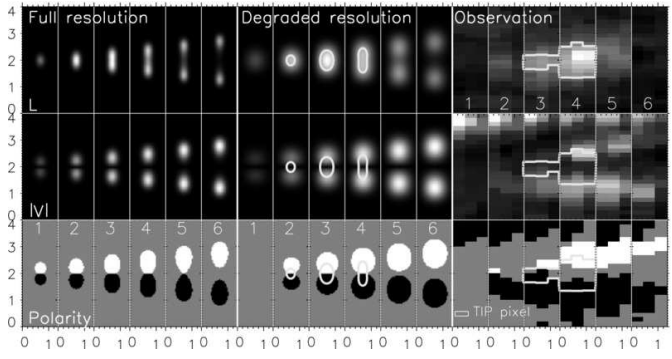}}}
         \caption{{\em Top to bottom}: Total linear polarization, 
                  circular polarization, and polarity. {\em Left}: 
                  Full resolution of the simulation. {\em Middle}: 
                  After degradation. {\em Right}: Corresponding 
                  section of the FoV in the observations. The 
                  {\em white rectangle} in the {\em bottom right 
                  panel} indicates the size of a pixel in the 
                  observations. {\em Contour lines} trace strong 
                  linear polarization signal. Each panel and 
                  quantity is displayed with an individual range 
                  for better visibility (see text for details). 
                  Tickmarks are in arcsec.
                 } 
          \label{2d_maps} 
  \end{figure}
%
      
\begin{figure}
          \resizebox{8.8cm}{!}{\includegraphics{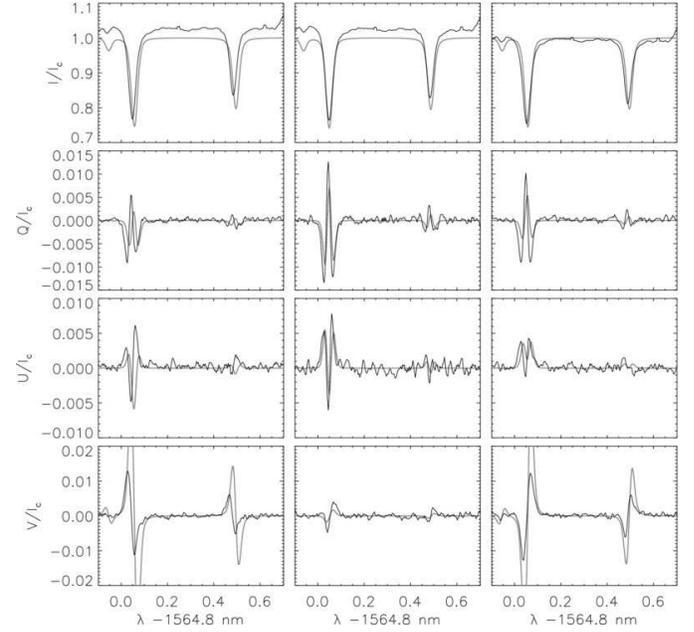}}
          \caption{Comparison of observed ({\em black}) and 
                   synthetic spectra ({\em gray}) in step 4. 
                   {\em Top to bottom}: Stokes $IQUV$. {\em 
                   Left to right}: Lower footpoint, center 
                   of loop, upper footpoint.
                   }
          \label{fig_profiles} 
   \end{figure}

At step 4 (10:42:10\,UT), the loop-like structure is seen most clearly 
in the observations 
(cf.\,Figs.~\ref{fig_gband_magparam} or \ref{2d_maps}). 
We thus selected the profiles at the center of the loop and both 
footpoints of this step from observations and the simulation ({\em short 
horizontal bars} in 
Fig.\,\ref{spec_comp}) 
for the direct comparison shown in 
Fig.\,\ref{fig_profiles}.
Besides the amplitude of Stokes $V$, the continuum intensity level and 
the Doppler shift in the lower footpoint, the match is surprisingly good, 
despite the simplifications used in generating the geometric model. We 
thus conclude that the observed spectra are in good agreement with a rising 
magnetic flux tube of around 250\,km diameter, 3$\times 10^{17}$\,Mx magnetic 
flux, and a rise speed of 1\,km\,s$^{-1}$.
%
   \begin{table}[!t]
          \caption{Display ranges for Fig.\,\ref{2d_maps}.}
          \label{disp_ranges_f11}
    \centering
      \begin{tabular}{l c c c}
        \hline
        \noalign{\smallskip}
          & simulation at   & convolved  & observations / dynamical \\
          & full resolution & simulation & range                  \\
        \noalign{\smallskip}
        \hline
        \noalign{\smallskip}
         $L$   & 0 - 1.18 & 0 - 0.4  & 1.23 - 1.71 / 0 - 0.48 \\
         $|V|$ & 0 - 4.1  & 0 - 1.48 & 0.9  - 1.6  / 0 - 0.7  \\
        \noalign{\smallskip}
        \hline
      \end{tabular}
  \end{table}
%
\section{Discussion and conclusions} \label{discus_and_concl}

High-resolution surveys of large areas of quiet-Sun internetwork 
(Lites et al. \cite{litesetal08}) 
and plage regions 
(Ishikawa et al. \cite{ishikawaetal08}) 
revealed widespread occurrences of relatively strong horizontal fields 
over granules, accompanied by vertical bipolar fields cospatial 
with up- and downflows 
(Mart\'{\i}nez Gonz\'{a}lez et al. \cite{martinezgonzalezetal07}). 
On the other hand, individual granules are often collocated with strong 
unipolar fields decoupled from granular outflows
(OR08, 
Bello Gonz\'{a}lez et al. \cite{bellogonzalezetal08}). 
%
%

CE07 
reported for the first time the emergence of a magnetic loop inside 
of a granule whose footpoints drifted towards intergranular lanes. 
Here we present new observational evidence for a similar event, 
whose onset manifests itself through enhanced net linear polarization 
within an evolved granule. Soon after, a pair of opposite circularly 
polarized patches appear symmetrically on each side of the linear 
polarization forming a small magnetic dipole bridged by a horizontal 
magnetic field. As long as both patches of circular polarization are 
visible, they are about of the same size and show a balanced magnetic 
flux of opposite sign ($+2.9$ and $-3.1\times 10^{17}$ Mx in step 4). 
This supports the conclusion made by 
Lamb et al. (\cite{lambetal08}) 
that the model of an asymmetric dipole emergence is very unlikely.

The fast disappearance of the linear polarization signal after 10:43:53\,UT 
(step 5, see Figs.\,\ref{fig_tip_overview}, \ref{fig_gband_magparam}, or \ref{2d_maps})
is in striking contrast with its gradual growth seen from 10:38:47\,UT 
to 10:43:53\,UT. Can we understand it just by an upward extending loop 
reaching up to a higher atmospheric layer undetectable by the 
\ion{Fe}{i} lines as suggested in 
CE07? 
If we consider the loop top as a magnetic perturbation propagating 
upwards through otherwise non-magnetic layers, then a smooth decay 
of contribution and response functions with height would imply slower 
ceasing of the net linear polarization, perhaps detectable on the 
next frame. Since the contribution and response functions of Stokes 
{\em Q} and {\em U} to magnetic perturbations can be very complex 
(del Toro Iniesta \cite{deltoroiniesta_book03}),
their detailed computations would help to understand the disappearance 
of linear polarization after 10:43:53\,UT. But after 10:43:53\,UT the 
circular polarization signal of the footpoint that meets an 
opposite-polarity magnetic field also weakens rapidly. This could indicate 
magnetic reconnection between the emergent flux and the preexisting 
magnetic fields. If reconnection thus is such a common feature, it 
could explain why OR08 found no trace of the flux emergence in the 
chromospheric layers.

The disappearance of the polarization signal could, however, also be 
related to other causes. In the rising loop simulation, the linear 
polarization amplitude at the last step is again comparable to the 
first one. Taking into account the noise present in the observations, 
the polarization signal could also be simply reduced below the detection 
limit. Another possibility could be a slight spatial drift of the 
feature with time that would remove it partly out of the observed FoV, 
because it already was located only in the first three steps of each 
repeated scan 
(see Fig.\,\ref{fig_tip_overview}). 
Moreover, the possible expansion of the magnetic loop with height 
(and time) could also affect the measurement of the linear signal above 
the detection threshold. From the comparison with the simulated loop, 
we conclude, however, that at least the ascent until maximal linear 
polarization signal is reached is fully traced by the observations.

Our flux density estimate of $\sim$450\,G found in the emerging loop 
is comparable to the values reported in 
Mart\'{\i}nez Gonz\'{a}lez et al. (\cite{martinezgonzalezetal07}) 
and
Ishikawa et al. (\cite{ishikawaetal08}) 
for small-scale magnetic loops in the solar internetwork and plage 
region, respectively; it significantly exceeds, however, the 
10-100\,G given by 
Mart\'{\i}nez Gonz\'{a}lez \& Bellot Rubio (\cite{martinezgonzalez_bellotrubio09}). 
This could be due to the assumption of a magnetic filling factor of 
unity by the latter authors. The loop footpoints separate with time 
by around 2\,km\,s$^{-1}$, drifting in opposite directions towards 
intergranular lanes. Moreover, our estimate of the magnetic flux density 
shows that the field is stronger than the typical equipartition value 
of 400\,G 
(Cheung et al. \cite{cheungetal07}, 
Ishikawa et al. \cite{ishikawaetal08}). 
This suggests that both granular outflow and relaxation of the magnetic 
tension of the loop may drive the footpoints.

How does our loop compare to the one reported in 
CE07? 
They occupy similar locations and enclose too little magnetic flux to 
leave any trace on the underlying granules or ambient G-band bright points. 
On the other hand, our loop displays a higher magnetic flux density and 
a longer lifetime. 
%
%

Another difference to the loop reported in 
CE07 
is that in our case one loop footpoint meets the ambient field of 
opposite polarity, which is followed by the fast disappearance of 
the loop, in contrast to its more gradual emergence. This is very 
symptomatic of field cancellation by magnetic reconnection, as  
reported in
Cheung et al (\cite{cheungetal08}) and 
Guglielmino et al. (\cite{guglielminoetal08}). 
The magnetic reconnection likely drags loop remnants very rapidly 
out of the line-forming domain less than 12 min after the first 
trace of the loop top was detected.

The magnetic loop emergence presented here is probably just a particular 
event in a multitude of similar ones occurring continuously in the solar 
internetwork. Simulations of small-scale magnetoconvection by, e.g.,  
Stein \& Nordlund (\cite{stein_nordlund06}) 
show examples of granular-sized magnetic loops with similar behavior. 
They probably build up a sub-canopy of horizontal fields over granules
(Wedemeyer-B\"{o}hm et al. \cite{wedemeyerbohmetal09}). 
Surface dynamo simulations by 
Sch\"{u}ssler \& V\"{o}gler (\cite{schussler_vogler08}) and 
Steiner et al. (\cite{steineretal08}) 
clearly show that this new component of the internetwork is a direct 
consequence of magnetic flux expulsion from the granular interior, 
not only to the intergranular lanes, but also to the upper photosphere. 
Cheung et al. (\cite{cheungetal07}, \cite{cheungetal08}) 
found in numerical simulations that emerging flux regions also 
contribute to the flux accumulation and its maintenance above granular 
upflows in the magnetic inversion layer. Our results show one particular 
example of this local process albeit terminated, probably due to magnetic 
reconnection. Better temporal and spatial resolution is needed for more 
detailed information on the process than our observations provide.

\begin{acknowledgements}

The VTT is operated by the Kiepenheuer-Institut f\"{u}r Sonnenphysik, 
Freiburg, at the Observatorio del Teide of the Instituto de 
Astrof\'{\i}sica de Canarias. The observations have been funded by the
OPTICON trans-national access program funded by the European Commission’s 
Sixth Framework Programme. The authors are grateful for the OPTICON 
support provided. Work of P.~G., J.~R., A.~K., and J.~K. was supported 
partly by the Slovak Research and Development Agency under the contract 
No. APVV-0066-06. This work also was partly supported by the Deut\-sche 
For\-schungs\-ge\-mein\-schaft grant No. DFG 436 SLK 113/13/0-1. We thank 
H.P.~Doerr (KIS) for the support with the Speckle computers at the VTT.
The authors would like to thank the anonymous referee for helpful 
comments. This research has made use of NASA's Astrophysics Data System.
\end{acknowledgements}



\begin{thebibliography}{}
   \bibitem[2007]{abbett07}
    Abbett, W.~P. 2007, 
      ApJ, 665, 1469

   \bibitem[2007]{becketal07}
    Beck, C., Mikurda, K., Bellot Rubio, L.R., Kentischer, T. \& 
    Collados, M. 2007, in Modern solar facilities - advanced solar science, 
    eds. F. Kneer, K.~G. Puschmann, \& A.~D. Wittmann, Proceedings of a Workshop 
    held at G\"{o}ttingen, September 27-29, 2006, G\"{o}ttingen, Germany, 
    Universit\"{a}tsverlag G\"{o}ttingen, p.\,55 (ISBN 978-3-938616-84-0)

   \bibitem[2005]{becketal05}
    Beck, C., Schlichenmaier, R., Collados, M., Bellot Rubio, L., 
      \& Kentischer, T.J. 2005, A\&A, 443, 1047

   \bibitem[2008]{bellogonzalezetal08}
    Bello Gonz\'{a}lez, N., Okunev, O., \& Kneer, F. 2008, 
      A\&A, 490, L23

   \bibitem[2004]{bellotrubioetal04}  
    Bellot Rubio, L.~R., Balthasar, H., \& Collados, M. 2004, 
      A\&A, 427, 319

   \bibitem[2006]{berkefeld06}
    Berkefeld, T. 2006, in Modern solar facilities - advanced solar science, 
    eds. F. Kneer, K.~G. Puschmann, \& A.~D. Wittmann, Proceedings of a Workshop 
    held at G\"{o}ttingen, September 27-29, 2006, G\"{o}ttingen, Germany, 
    Universit\"{a}tsverlag G\"{o}ttingen, p.\,107 (ISBN 978-3-938616-84-0)

   \bibitem[2005]{cabrera_solanaetal05}
    Cabrera Solana, D., Bellot Rubio, L.~R., \& del Toro Iniesta, J.~C. 2005, 
      A\&A, 439, 687

   \bibitem[1999]{cattaneo99}
    Cattaneo, F. 1999, 
      ApJ, 515, L39

   \bibitem[2007]{centenoetal07}
    Centeno, R., Socas-Navarro, H., Lites, B., \& et al. 2007, 
      ApJ, 666, L137 (CE07)

   \bibitem[2007]{cheungetal07}
    Cheung, M.~C.~M., Sch\"{u}ssler, M., \& Moreno-Insertis, F. 2007, 
      A\&A, 467, 703
 
   \bibitem[2008]{cheungetal08}
    Cheung, M.~C.~M., Sch\"{u}ssler, M., Tarbell, T.~D., \& Title, A.~M.
      2008, ApJ, 687, 1373

   \bibitem[1999]{collados99}
    Collados, M. 1999, in Third Advances in Solar Physics Euroconference:
    Magnetic Fields and Oscillations, eds. B. Schmieder, A. Hofmann, \& 
    J. Staude, ASP Conference Series 184, p.\,3 (ISBN 1-58381-010-2)

   \bibitem[2003]{collados03} 
    Collados, M. 2003, in Polarimetry in Astronomy, ed. S. Fineschi, 
    Proceedings of the SPIE -  The International Society for Optical 
    Engineering, vol. 4843, p.\,55

   \bibitem[2007]{colladosetal07}
    Collados, M., Lagg, A., D\'{\i}az Garc\'{\i}a, J.~J., \& et al. 2007, 
    in The Physics of Chromospheric Plasmas, eds. P. Heinzel, I. 
    Dorotovi\v{c}, \& R.~J. Rutten, Proceedings of the conference held at 
    the University of Coimbra, October 9-13, 2006, Coimbra, Portugal, 
    ASP Conference Series 368, p.\,611

   \bibitem[1990]{georgakilasetal90}
    Georgakilas, A.~A., Alissandrakis, C.~E. \& Zachariadis, T.~G. 1990, 
      Solar Physics, 129, 277

   \bibitem[1971]{gingerichetal71}
    Gingerich, O., Noyes, R.~W., Kalkofen, W., \& Cuny, Y. 1971, 
      Solar Physics, 18, 347

   \bibitem[2008]{guglielminoetal08} 
    Guglielmino, S.~L., Zuccarello, F., Romano, P., \& Bellot Rubio, L.~R.  
      2008, ApJ, 688, L111

  \bibitem[2008]{ishikawaetal08}
    Ishikawa, R., Tsuneta, S., Ichimoto, K., \& et al. 2008, 
      A\&A, 481, L25
 
   \bibitem[1998]{kentischeretal98}
    Kentischer, T.~J., Schmidt, W., Sigwarth, M., \& Uexk\"{u}ll, M.~V.
      1998, A\&A, 340, 569

   \bibitem[2007]{khomenko_collados07}
    Khomenko, E., \& Collados, M. 2007, 
      ApJ, 659, 1726

   \bibitem[2007]{kosugietal07}
    Kosugi, T., Matsuzaki, K., Sakao, T., \& et al. 2007, 
      Solar Physics, 243, 3
 
   \bibitem[2008]{kuceraetal08}
    Ku\v{c}era, A., Beck, C., G\"{o}m\"{o}ry, P., \& et al. 2008, 
    in Proceedings of the 12$^{th}$ European Solar Physics Meeting, September 8-12, 
    2008, Freiburg, Germany. p.\,2.52.~Available online at 
    http://espm.kis.uni-freiburg.de.  
    
   \bibitem[2008]{lambetal08}
    Lamb, D.~A., DeForest, C.~E., Hagenaar, H.~J., Parnell, C.~E., \& 
      Welsch, B.~T. 2008, ApJ, 674, 520

   \bibitem[2001]{litesetal01}
    Lites, B.~W., Elmore, D.~F.,\& Streander, K.~V. 2001, 
    in Advanced Solar Polarimetry - Theory, Observation, and Instrumentation,
    ed. M. Sigwarth, Proceedings of the 20th NSO/Sac Summer Workshop, 
    ASP Conference Series 236, p.\,33 

   \bibitem[2008]{litesetal08}
    Lites, B.~W., Kubo, M., Socas-Navarro, H., \& et al. 2008, 
      ApJ, 672, 1237

   \bibitem[2003]{vdlueheetal03}
    von der L{\"u}he, O., Soltau, D., Berkefeld, T., \& Schelenz, T. 2003, 
    Proceedings of the SPIE, 4853, 187, ed.~by S.~L. Keil, \& S.~V. 
    Avakyan

   \bibitem[2009]{martinezgonzalez_bellotrubio09}
    Mart\'{\i}nez Gonz\'{a}lez, M.~J., \& Bellot Rubio, L.~R. 2009, 
      ApJ, 700, 1391

   \bibitem[2007]{martinezgonzalezetal07}
    Mart\'{\i}nez Gonz\'{a}lez, M.~J., Collados, M., Ruiz Cobo, B.,
     \& Solanki, S.~K. 2007, A\&A, 469, L39

   \bibitem[1996]{meinetal96}
    Mein, P., Demoulin, P., Mein, N., \& et al. 1996, 
      A\&A, 305, 343

   \bibitem[2008]{orozcosuarezetal08} 
    Orozco Su\'{a}rez, D., Bellot Rubio, L.~R., del Toro Iniesta, J.~C.,
      \& Tsuneta, S. 2008, A\&A, 481, L33 (OR08)

   \bibitem[1992]{cobo_iniesta92} 
    Ruiz Cobo, B., \& del Toro Iniesta, J.~C. 1992,
      ApJ, 398, 375

   \bibitem[2006]{schaffenbergeretal06}
    Schaffenberger, W., Wedemeyer-B\"{o}hm, S., Steiner, O., \& 
    Freytag, B. 2006, in Solar MHD Theory and Observations: A High Spatial 
    Resolution Perspective, eds. J. Leibacher, R.~F. Stein, \& H. Uitenbroek, 
    Proceedings of the conference held at the National Solar Observatory, 
    July 18-22, 2005, Sacramento Peak, New Mexico, USA, 
    ASP Conference Series 354, p.\,345  

   \bibitem[2002]{schlichenmaier_collados02}
    Schlichenmaier, R., \& Collados, M. 2002, 
      A\&A, 381, 668

   \bibitem[1985]{schroteretal85} 
    Schr\"{o}ter, E.~H., Soltau, D., \& Wiehr, E. 1985,	
      Vistas in Astronomy, 28, 519

   \bibitem[2008]{schussler_vogler08}
    Sch\"{u}ssler, M., \& V\"{o}gler, A. 2008, 
      A\&A, 481, L5

   \bibitem[2006]{stein_nordlund06}
    Stein, R.~F., \& Nordlund, \AA. 2006, 
      ApJ, 642, 1246
 
   \bibitem[2008]{steineretal08}
    Steiner, O., Rezaei, R., Schaffenberger, W., \& Wedemeyer-B\"{o}hm, S.
      2008, ApJ, 680, L85

   \bibitem[2003]{deltoroiniesta_book03}
    del Toro Iniesta, J.~C. 2003, 
    Introduction to Spectropolarimetry, ISBN 0521818273. Cambridge, UK: 
    Cambridge University Press

   \bibitem[2002]{tritschleretal02}
    Tritschler, A., Schmidt, W., Langhans, K., \& Kentischer, T. 2002, 
      Solar Physics, 211, 17

   \bibitem[2005]{vogleretal05}
    V\"{o}gler, A., Shelyag, S., Sch\"{u}ssler, M., \& et al. 2005, 
      A\&A, 429, 335

   \bibitem[2009]{wedemeyerbohmetal09}
    Wedemeyer-B\"{o}hm, S., Lagg, A., \& Nordlund, \AA. 2009, 
      Space Science Reviews, 144, 317

   \bibitem[2008]{wogeretal08}
    W\"{o}ger, F., von der L\"{u}he, O., \& Reardon, K. 2008, 
      A\&A, 488, 375
\end{thebibliography}
\end{document}